\documentclass[aps,prb,twocolumn,groupedaddress,citeautoscript]{revtex4}

\usepackage{graphicx}
\usepackage{amsmath}
\usepackage{times}

\graphicspath{{.}{./EPS/}}

\begin{document}

\title{The Quantized Hall Insulator:
a ``Quantum" Signature of a ``Classical" Transport Regime?}

\author{Efrat Shimshoni}
\affiliation{Department of Mathematics--Physics, 
University of Haifa at Oranim, Tivon 36006, Israel}

\date{\today}

\begin{abstract}

Experimental studies of the transitions from a primary quantum Hall (QH) 
liquid at filling factor
$\nu=1/k$ (with $k$ an odd integer) to the insulator have indicated
a ``quantized Hall insulator'' (QHI) behavior: while the longitudinal 
resistivity diverges with decreasing temperature and current bias, the Hall
resistivity remains quantized at the value $k h/ e^2$. We review the 
experimental results and the theoretical studies addressing this phenomenon.
In particular, we discuss a theoretical approach which employs a model of
the insulator as a random network of weakly coupled puddles of QH liquid at 
fixed $\nu$. This model is proved to exhibit a robust quantization 
of the Hall resistivity, provided the electron transport on the network is 
{\em incoherent\/}. Subsequent theoretical studies have focused on the 
controversy whether the assumption of incoherence is necessary. The emergent
conclusion is that in the quantum coherent transport regime, 
quantum interference destroys the QHI as a consequence of localization. 
Once the localization length becomes much shorter than the dephasing 
length, the Hall resistivity diverges. We conclude by mentioning some recent 
experimental observations and open questions.

\end{abstract}

\maketitle

\section{Introduction}
\subsection{The concept of ``Hall Insulator''}
An electric insulator is defined as a state of the electronic system where
charge transport is strongly suppressed. In an ideal insulator, an infinite 
voltage is required in order to pass current through the system. However, 
physical insulators are studied experimentally in conditions where the 
temperature ($T$), the frequency of the driving source ($\omega$) and the
voltage bias ($V$) are finite, and typically do not exhibit an ideal insulating
behavior. It is nevertheless possible to classify a quantum many--body 
electronic state as an ``insulator'', provided its resistance for current 
flowing along the direction of an applied voltage diverges in the limit
where $T,\omega,V\rightarrow 0$ and the system size is infinite.

In the presence of a magnetic field ${\bf B}=B {\bf \hat z}$, the conductivity 
tensor acquires an off-diagonal component $\sigma_{xy}$. An insulating 
behavior is then characterized by the vanishing of 
$\sigma_{xx}$ and $\sigma_{xy}$ and correspondingly 
a divergence of the longitudinal resistivity $\rho_{xx}$, which 
distinguishes it quite unambiguously from a conducting state. In contrast, 
the {\em Hall resistance}
\begin{equation}
\rho_{xy}=\frac{\sigma_{yx}}{(\sigma_{xx}^2+\sigma_{xy}^2)}
\end{equation}
is a subtle quantity: its behavior in the limit $T,\omega,V\rightarrow 0$
depends crucially on the ratio $\sigma_{xy}/\sigma_{xx}$ when both 
components approach zero. In particular, in the case where the scaling
relation $\sigma_{xy}\sim\sigma^2_{xx}$ holds, $\rho_{xy}$ is {\em finite}.
Such a peculiar insulating state, in which the Hall resistance is
essentially indistinguishable from a conductor, is named a ``Hall Insulator''.

A number of theoretical studies\cite{Fuku,Efetov,KLZlett,Joe,Herb} have 
argued that a Hall insulating behavior, where $\rho_{xy}\sim B$ as in a Drude 
conductor, is actually a quite generic property of disordered non--interacting
electron systems. These correspond to electronic states where the insulating
character is a consequence of Anderson localization; the behavior of 
$\rho_{xy}$ possibly marks their distinction from band insulators or
Mott (interaction--dominated) insulators. However, the dominant role of
disorder in this case poses a difficulty: a theoretical evaluation of the
transport coefficients involves a disorder averaging procedure, and is hence
quite subtle. Indeed, the existence of a Hall Insulator was challenged by 
Entin--Wohlman {\it et al.}\cite{Ora}, who showed that a direct derivation of
the resistivity tensor (as opposed to inversion of a calculated conductivity 
tensor) yields an exponentially divergent $\rho_{xy}$ for $T\rightarrow 0$. 
They further argue that this behavior is to be expected in a ``true d.c.''
measurement. Experimental studies fail to settle this controversy: while some
data\cite{FrPo} support the claims of Ref. [6], other experiments
have reported an observation of a Hall Insulator\cite{Hop}. 

\subsection{The Insulating Phase in the Quantum Hall Regime}
The behavior of the Hall resistance is of particular significance in the so 
called ``quantum Hall (QH) regime'', which characterizes the low--$T$ 
behavior of disordered two--dimensional (2D) electron systems 
subject to a strong perpendicular magnetic field. These systems exhibit a rich
phase diagram indicating a multitude of transitions 
between phases with distinct transport properties\cite{qhe-phas-tran}. These 
include primarily the various QH phases, characterized by 
quantized values of the Hall resistivity: $\rho_{xy}=h/e^2 
\nu$ in a wide range of carrier densities $n$ and magnetic fields $B$ 
centered around certain rational values of the filling factor 
$\nu=n\phi_0/B$ (where $\phi_0=hc/e$ is the flux quantum). These plateaus in 
$\rho_{xy}$ are accompanied by a {\em vanishing} longitudinal resistivity 
$\rho_{xx}$. At sufficiently strong magnetic field or disorder, the 
series of QH--to--QH transitions is terminated by a transition to 
an insulator, marked by a {\em divergence\/} of $\rho_{xx}$.

Based on the flux--attachment mapping of electrons in the QH regime to
``composite (Chern--Simons) bosons'', and a resulting set of laws of 
corresponding states, Kivelson, Lee and Zhang (KLZ)\cite{KLZ} have argued that
the above mentioned insulating phase should exhibit the Hall Insulator 
behavior $\rho_{xy}\sim B/nec$. Their theory, however, does not provide
a clean derivation of this expression for $\rho_{xy}$ except at the particular
quantized values of $\nu$. The classical, Drude--like linear dependence on $B$
is therefore presented as a suggestive interpolation. This behavior is 
consistent with the experimental data of Goldman {\it et al.}\cite{Vladimir},
in apparent contradiction with the exponential divergence of $\rho_{xy}$
reported by Willett {\it et al.}\cite{Willett} at low $\nu$. The latter, 
however, can possibly be interpreted as evidence for a Wigner crystal, which
is expected to form at sufficiently low filling factors. These studies 
indicate that the insulating state in a strong magnetic field is not 
necessarily unique: the Hall resistance may serve as a probe which effectively 
distinguishes between different mechanisms for an insulating behavior.

\section{Experimental Evidence for a Quantized Hall Insulator}

\begin{figure}[b]
\includegraphics[width=2.7in]{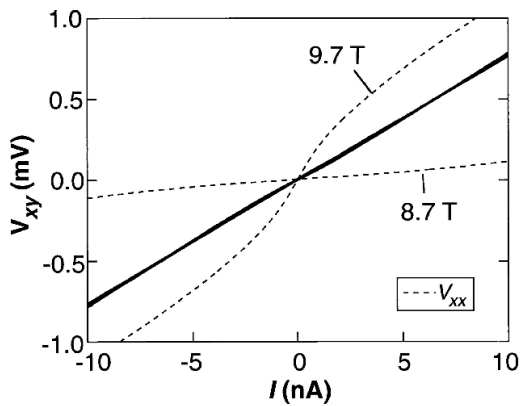}
\vspace*{8pt}
\caption{Plots of the Hall voltage $V_{xy}$ versus the current $I$
taken at 0.1-T steps in the $B$ range from 8.7 T to 9.7 T (the transition from 
$1/3$--QH state to the insulator occurs at $B_c=9.1$). Dashed traces are
of $V_{xx}$. $T=21$mK. [Figure taken from Ref. 14].}
\label{exp-3-I}
\end{figure}

\begin{figure}[b]
\includegraphics[width=2.7in]{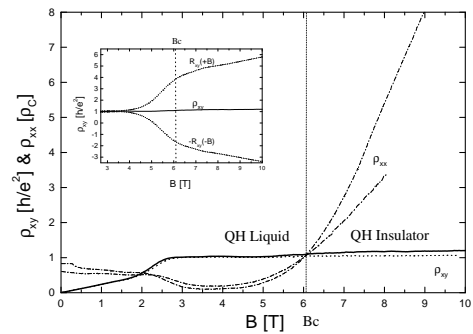}
\vspace*{8pt}
\caption{The Hall and diagonal resistivities as a function of 
$B$-field. The solid line is the Hall resistivity measured at 
T=300 mK and a current I=200 nA, whereas the dotted line 
is for I=400 nA. 
The dash-dotted lines are $\rho_{xx}$ at T=1.2 K (the 
uppermost curve) and at T=4.2 K (the lower curve). $V_G$=5.2 V, 
$\rho_c=1.65 h/e^2$ and $B_C$=6.06 T.
The inset shows the Hall resistances for $+B$ and $-B$ in dotted lines 
and $\rho_{xy}$ with a solid line. [Figure taken from Ref. 17].}
\label{exp-1-I}
\end{figure}

Significant insight on the nature of the insulating phase in the QH regime 
was gained by later experimental studies of the transition from the primary
QH states $\nu=1/k$ (where $k$ is an odd integer) to the neighboring 
insulator, driven by an increasing magnetic field $B$. In particular, 
a striking resemblance of the transport properties in the vicinity of
the transition to the behavior near a superconductor--to--insulator 
transition in thin films\cite{Shahar1} have inspired a further investigation
of these transitions in view of the theoretical framework of bosonic 
Chern-Simons used by KLZ, which essentially mapps the QH states to 
superconducting states of the composite bosons. Most remarkably, Shahar 
{\it et al.}\cite{SSSexp} have found that the longitudinal
current--voltage characteristics $I_x(V_{xx})$
in the QH and insulating phases are related by a ``reflection" symmetry,
i.e. a symmetry to trading the roles of current and voltage. It was 
shown\cite{SSSexp,SSSthe} that this symmetry can be interpreted as 
evidence for charge-flux duality relating the charged composite bosons 
in one phase to vortices in the other. This interpretation relies on the
mapping of the observable resistivity tensor $\rho_{ij}$ to the fictitious
resistivity tensor of the composite bosons, $\rho^b_{ij}$: 
\begin{eqnarray}
\rho_{xx} &=& \rho^b_{xx}\; ,\nonumber \\
\rho_{xy} &=& \rho^b_{xy}+k\frac{h}{e^2}\; ,\label{reslaw}
\end{eqnarray}
a relation which has been argued to be valid beyond linear 
response\cite{SSSthe}.

By inspection of the ``resistivity law" [Eq. (\ref{reslaw})] it is easy 
to see that the reflection symmetry 
\begin{equation}
\rho_{xx}(\nu)=1/\rho_{xx}(\nu_d) \label{refsym}
\end{equation}
(where $\nu$ and $\nu_d$ are filling factors in the QH and insulating 
phases, respectively, and units where $h/e^2=1$ are used) 
is unambiguously equivalent to the duality
relation in the composite bosons representation
\begin{equation}
\rho^b_{ij}(\nu)=\sigma^b_{ji}(\nu_d) \label{dual}
\end{equation}
only provided $\rho^b_{ij}$ is diagonal, i.e. $\rho^b_{xy}=0$. Eq. 
(\ref{reslaw}) then leads to the unavoidable conclusion that
the observable Hall resistivity should be $\rho_{xy}=kh/e^2$ 
{\em in both the QH and insulating phases}. Thus, based on 
``circumstantial evidence", it has been predicted that the 
Hall resistivity in the insulator neighboring a particular $1/k$--QH 
state should remain {\em quantized} at the plateau value. This 
intriguing behavior named a ``quantized Hall insulator" (QHI) was indeed 
confirmed experimentally by Shahar {\it et al.}\cite{SSSexp}, who studied
the transition from a $1/3$--QH state to the insulator in a moderately 
disordered GaAs sample. As indicated in Fig. \ref{exp-3-I},
all traces of $V_{xy}$ vs. $I$, including those taken in the insulator 
($B>B_c$ where $B_c$ is the critical field), collapse on a single, 
linear trace with a quantized slope: $\rho_{xy}=V_{xy}/I=3h/e^2$. 
Moreover, the quantization in the Hall response has been observed in a regime
where the longitudinal current--voltage relation is non--linear.

Subsequent experimental studies of $\rho_{xy}$ in the vicinity of other 
QH transitions, and in different types of samples and materials,
have confirmed this observation\cite{qhi-expA,qhi-expB,qhi-expC}. For example,
Fig. \ref{exp-1-I} presents the transition from the integer 
$\nu=1$ QH phase to the 
insulator in a two--dimensional (2D) hole system confined in a Ge/SiGe 
quantum well. Note that to eliminate the mixing with the longitudinal 
component of the resistivity (which is particularly large in the insulator), 
the Hall resistance is measured in two opposite orientations of the $B$; 
$\rho_{xy}$ is defined as the antisymmetric component: 
$\rho_{xy}=1/2[R_{xy}(+B)-R_{xy}(-B)]$ (see inset of Fig. \ref{exp-1-I}). 

The accumulated experimental evidence for the existence of QHI states
raised a number of questions. Most prominently, when combined with
earlier experimental results\cite{Vladimir,Willett} it appears to 
support the possibility that the insulating regime in the global QH 
phase diagram does not consist of a single type of insulator. Rather,
it includes a sequence of QHI states, characterized by different values 
of the odd integer $k$, and in between regions where $\rho_{xy}$ is not
quantized, and exhibits either the classical behavior $\rho_{xy}\sim B$
or diverges. It is not clear, however, whether these are different
insulating {\em phases} separated by critical manifolds in the
phase diagram, or different transport regimes which connect to each
other by a smooth cross--over. In spite of the valuable insight gained
by the bosonic Chern--Simons theory and its role in the initial discovery 
of the QHI, it does not seem to provide tools for a more detailed study
of the transport which is required to address these issues. An alternative
theoretical framework, which proved to be more practical, is reviewed
in the next Section.

\section{Theoretical Models}
The physics of the integer QH effect is described quite well in terms of
the Landau level quantization of non--interacting electrons. In particular,
in very strong $B$ such that only the first Landau level is partially occupied,
and in the presence of a slowly varying disorder potential, one can use
a semiclassical approximation\cite{Trugman} in which the electrons at the
Fermi level travel along equipotential contours. Scattering among 
equipotential contours is possible in the vicinity of saddle points in the
potential via tunneling. As a result, transport in the 2D systems is 
effectively carried by a network of quasi one--dimensional 
chiral channels (edge states), and can 
be modeled within the Landauer--B{\"u}ttiker approach\cite{LanBut}.
This provides a convenient framework for the study of a transition from a
$\nu=1$ QH state to the insulator, where the two phases are defined according
to the behavior of the transmission through the sample (${\cal T}$)
in the $T\rightarrow 0$ limit: ${\cal T}\rightarrow 1$ in the QH phase,
${\cal T}\rightarrow 0$ in the insulator. This description can be generalized
to any of the transitions from a $1/k$--QH state to the insulator using a 
mapping to composite fermions\cite{SSSthe}. As we show below, both the 
``reflection symmetry'' of $\rho_{xx}$ and the behavior of $\rho_{xy}$ can be
interpreted more intuitively in this approach.

\subsection{The Single Scatterer: a Model for the Two--Terminal Configuration}
\begin{figure}[b]
\includegraphics[width=2.7in]{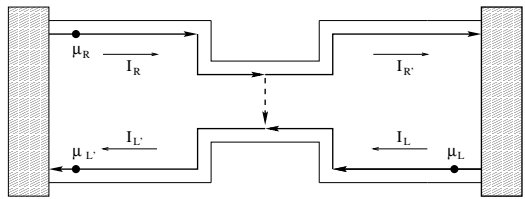}
\vspace*{8pt}
\caption{A schematic illustration of a Hall bar connected to
two reservoirs (shaded regions).}
\label{landauer}
\end{figure}

The robust quantization of $\rho_{xy}$ was demonstrated by Jain and 
Kivelson\cite{JK} in a QH system assuming a single saddle--point in the 
potential. They consider a Hall bar connected between two terminals, as 
depicted in Fig. \ref{landauer}, where the single constriction (schematically
representing the saddle--point) is characterized by a transmission probability
${\cal T}$. In general, ${\cal T}$ depends on the potential, the filling factor
(where here $0<\nu<1$) and possibly (in the non--linear response regime)
on the current bias $I=I_R-I_{L'}$. The currents in Fig. \ref{landauer} are 
related by
\begin{equation}
I_{L'}={\cal T}I_L+{\cal R}I_R\; ,
\end{equation}  
where ${\cal R}=1-{\cal T}$ is the reflection probability. 

Away from the
constriction there is no scattering, and a local equilibrium is assumed so
that the chemical potential and the current are related by\cite{Been}
\begin{equation}
\mu_i=\frac{h}{e^2}I_i\; ,\quad {\rm where}\quad i=R,R',L,L'\; .
\label{muI}
\end{equation}
The longitudinal voltage measured between the two terminals is then given by
\begin{equation}
V_{xx}=\mu_{L'}-\mu_L=\frac{{\cal R}}{{\cal T}}\frac{h}{e^2}I\; .
\label{Vxx}
\end{equation}
At the same time, the Hall voltage is given by
\begin{equation}
V_{xy}=\mu_R-\mu_{L'}=\mbox{sgn}(B) \frac{h}{e^2}I\; .
\label{Vxy}
\end{equation}
The factor $\mbox{sgn}(B)$ accounts for the fact that
a reversal of $B$ for a fixed current $I$ (which correspond to
reversing the direction of local currents $I_i$ and redefining the current
as $I=I_{L'}-I_R$) reverses $V_{xy}$ [i.e. $V_{xy}(B)=-V_{xy}(-B)$], while
leaving $V_{xx}$ unchanged.
This implies that although the voltage measured between arbitrary two points
across the Hall bar may include a longitudinal component (e.g., 
$\mu_R-\mu_L=V_{xx}+V_{xy}$), the antisymmetrization with respect to 
$B\rightarrow -B$ eliminates $V_{xx}$, yielding the pure Hall resistance
$\rho_{xy}=V_{xy}/I=h/e^2$.

Two points should be emphasized with regards to the above result. First,
$\rho_{xy}$ is quantized regardless of any details of the scatterer:
the dependence on the potential, magnetic field and bias affect
the pure longitudinal response alone. In particular, the quantization survives
in the insulating regime corresponding to ${\cal T}\rightarrow 0$,
where $\rho_{xx}\rightarrow \infty$. The second interesting point is that
an exchange of ${\cal T}$ and ${\cal R}$ (a procedure which corresponds to
the trading of particles and holes compared to the half--filled Landau level),
maps $\rho_{xx}$ on $1/\rho_{xx}$ [see Eq. (\ref{Vxx})]. The charge--flux 
duality, which has been argued to be responsible for the reflection 
symmetry within the bosonic Chern--Simons theory, is therefore equivalent
to a particle--hole symmetry in the ``fermionic'' picture\cite{SSSthe}. 

The above analysis can be easily generalized to the more realistic
case of a macroscopic QH system, where the disorder potential includes a 
multitude of saddle--points with random parameters. It is still possible
to define an overall transmission probability across the Hall bar connected
between the terminals. This suggests, that the single--scatterer model provides
a complete transport theory, which explains both the reflection symmetry and
the QHI phenomenon quite directly. As a matter of fact, the latter appears
to be rather trivial! However, one should keep in mind that the measurement 
configuration depicted in Fig. \ref{landauer} is not an ideal setup for
the probing of the resistivity tensor in a 2D Hall bar. The system is
connected between two terminals only, each corresponding to a single pair  
of outgoing and incoming conducting channels. The Hall voltage as defined
in such a connection (where the Hall probes are attached to the very
same outgoing and incoming channels), is indeed trivially related to the 
current. In contrast, as we show in the next Sections, an ``honest" 
measurement of $V_{xy}$ which involves {\em four} independent contacts, 
is more informative. In particular, it does not necessarily yield a 
quantized $\rho_{xy}$.

\subsection{The ``Classical" Puddle--Network Model}
The behavior of a most general integer QH system at $0<\nu<1$ is
best captured by network models, e.g. of the type proposed by 
Chalker and Coddington (CC)\cite{CC}. The sample is represented by a network of
QH puddles encircled by chiral edge channels, which are connected to each other
by constrictions of the type depicted in Fig. \ref{landauer}, each 
characterized by transmission and reflection amplitudes $t$ and $r$. The 
latter are assumed to be random variables. 

A study of the properties of local currents in such a network 
model\cite{ruzin} leads to the derivation of a ``semi--circle" law 
for the conductivities in the vicinity of a transition from the 
$n+1$ to the $n$ QH state
\begin{equation}
\sigma_{xx}^2+\left[\sigma_{xy}-(n+1/2)\frac{e^2}{h}\right]^2
=\left[\frac{1}{2}\frac{e^2}{h}\right]^2\; ,
\label{semicirc}
\end{equation}
which in the special case $n=0$ ($(\nu=1)$--QH state to the insulator) 
is essentially equivalent to the statement that $\rho_{xy}$ is 
quantized. This derivation is restricted to linear response and relies
on the existence of a local conductivity tensor. The quantization of the
Hall resistance was proved to be more robust, and in particular
valid beyond linear response, in a later work\cite{SA} employing the
Landauer approach to transport which allows a direct derivation of the
resistivity rather than conductivity tensor. Below we sketch this derivation.

\begin{figure}[b]
\includegraphics[width=2.7in]{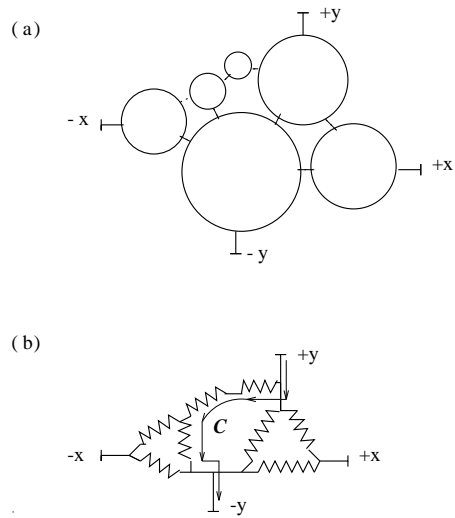}
\vspace*{8pt}
\caption{(a) Typical puddle network, where dotted lines represent 
constrictions. (b) Corresponding equivalent circuit. 
Path ${\cal C}$ is denoted by arrows. [Figure taken from Ref. 25].}
\label{pudnet}
\end{figure}

Consider a general network of QH puddles at filling factor $1/k$ (with $k$ an
odd integer) separated by constrictions (tunnel barriers), 
with two current leads at $-x$ and $+x$, 
and two voltage leads at $-y$ and $+y$ [see Fig. \ref{pudnet}(a)]. 
Assuming that the typical distance between adjacent constrictions is
sufficiently large to establish local equilibrium, it is possible to
define for each constriction connecting puddle $i$ to $j$ a local Hall voltage 
\begin{equation}
V_{xy}(ij)=\mbox{sgn}(B) \frac{h}{e^2} k I_{ij}\; ,
\end{equation} 
where $I_{ij}$ is the current flowing from $i$ to $j$. This is essentially a
generalization of Eq. (\ref{Vxy}) to the fractional QH case $k\not=1$. In
addition, it is assumed that all quantum interference effects take place 
within the tunnel barrier length scales, beyond which dephasing effects 
destroy coherence between tunneling events. As a consequence,
the longitudinal response of the network can be modeled by the equivalent 
circuit depicted in Fig. \ref{pudnet}(b). Namely, it behaves as a classical 
2D resistor network in which vertices represent the QH puddles, and the 
resistors mimic the constrictions between them, each
characterized by a local current--voltage relation $V_{ij}(I_{ij})$.
For a given total current $I$ driven between $-x$ and $+x$, the distributions 
of $\{V_{ij}\}$ and $\{I_{ij}\}$ in the network is dictated by the classical 
Kirchoff's laws. It can be proved\cite{SA}, that the total number of equations
imposed by this set of constraints is exactly sufficient to determine 
$\{I_{ij}\}$ uniquely. 

The total transverse voltage $V_y$ can be evaluated by choosing any path  
${\cal C}$ which connects the $-y$ lead to the $+y$ lead (see Fig. 
\ref{pudnet}). Summing up all contributions one obtains
\begin{equation}
V_y(B)= \sum_{(ij)^L\in{\cal C}} V_{ij}(I_{ij},|B|) +\mbox{sgn}(B) k 
\frac{h}{e^2}\sum_{(ij)^H\in{\cal C}} I_{ij}\; ,
\label{Vytotal}
\end{equation} 
where  $(ij)^L$ denote sections of the path along the resistors, and $(ij)^H$
the sections connecting two points across vertices in the graph. 
Here we account for the fact
that $V_{ij}$ typically depend on $B$, but obeys $V_{ij}(B)=V_{ij}(-B)$.
Eq. (\ref{Vytotal}) indicates quite transparently that the longitudinal
and Hall components of the voltage drop between any two points 
are conveniently separated by symmetry under reversal of $\mbox{sgn}(B)$.
In particular, defining the Hall voltage to be the antisymmetric component 
$V_{xy}=1/2[V_y(B)-V_y(-B)]$, it is easy to see that the contribution from
the first term in the expression for $V_y$ cancels out. Since by global 
current conservation the second term is proportional to the total current $I$,
one obtains
\begin{equation}
V_{xy}=\mbox{sgn}(B) k \frac{h}{e^2}I\; .
\label{VxyQHI}
\end{equation}
The Hall resistivity is therefore quantized at $\rho_{xy}= k (h/ e^2)$, and
{\em completely independent} of $B$ and $I$.

The above model for the transport implies that a quantized Hall resistivity
is a remarkably robust feature of any QH system in both sides of the 
transition from a $1/k$ QH liquid to the insulator. All details of the 
disorder potential, the dependence on filling factor and temperature as well as
deviations from linear response are reflected by the longitudinal response 
alone. In particular, the quantization persists in both the QH liquid and 
insulating phases, in agreement with the experimental 
observations\cite{SSSexp,qhi-expA,qhi-expB,qhi-expC}. However, at this point 
it should be emphasized that the proof (which incorporates the use of
the classical Kirchoff's laws) relies on a crucial assumption: that there is
a strong dephasing mechanism, responsible for the suppression of quantum
interference anywhere except the close vicinity of the tunnel barriers between
adjacent puddles. This scenario is valid as long as the typical puddle--size 
$L_p$ is large compared to the dephasing length $L_\phi$. 
The condition $L_p>L_\phi$ is expected to be violated, 
either at sufficiently low $T$ (where $L_\phi$ diverges), or
due to the reduction of puddle--size sufficiently far from the transition to 
the insulator (or in the presence of a short--range impurity potential). The
question arises, whether the requirement of complete dephasing is indeed
necessary. As discussed in the next subsection, it turns out that answering 
this question is not straightforward.

\subsection{The Role of Quantum Interference: Breakdown of the QHI}
In the purely quantum, phase--coherent limit, electron transport in 
a network of the type depicted in Fig. \ref{pudnet}(a) is no longer 
dictated by Kirchoff's laws. Local currents in the network are given 
by summing current {\em amplitudes} rather than current 
{\em probabilities}. Such a coherent summation introduces interference 
terms which are likely to have dramatic influence on the transport. In 
particular, interference effects are known to be responsible for the 
emergence of localization. Obviously, the transport problem becomes 
far less tractable in this limit. Indeed,
attempts to establish a theoretical description of the transport 
coefficients, and the behavior of the Hall resistance in particular,
were subject to controversies over the last few years.

The quantum limit was first addressed by Ruzin and Feng\cite{ruzin}, 
who argued that the semi--circle law Eq. (\ref{semicirc}) (and hence,
in particular, the quantization of $\rho_{xy}$ in the insulator
neighboring a QH state) is still valid in this limit. Their derivation
employs a description of the QH system in terms of the chess-board-like 
(CC) version of the network model\cite{CC}. Defining
the local currents in adjacent patches (representing the QH states $n$ and 
$n+1$) to be $J_1$ and $J_2$, respectively, their proof of Eq. 
(\ref{semicirc}) proceeds in two stages. First, they prove the equivalence
of the semi--circle law to the statement that $J_1$ and $J_2$ are 
orthogonal to each other. Then, they demonstrate the orthogonality of
the currents. The latter property is shown to be valid in the quantum
limit as well, using a numerical computation. However, the proof of the first
part relies on the assumption of local equilibrium. Hence, the statement
of equivalence to the semi--circle law is not clearly
applicable in the quantum coherent limit.

\begin{figure}[b]
\includegraphics[width=2.7in]{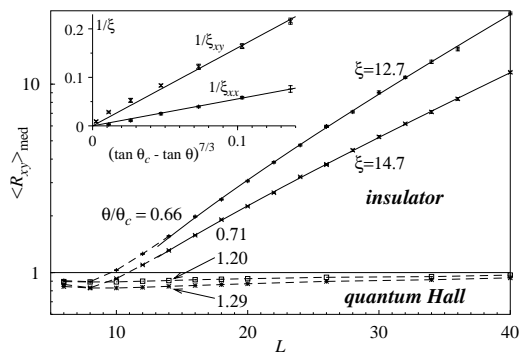}
\vspace*{8pt}
\caption{Finite size scaling of the Hall resistance in the QH and insulator
phases. $\xi$ denotes the localization length, where $\xi_{xx}$ and 
$\xi_{xy}$ correspond to $\rho_{xx}$ and $\rho_{xy}$, respectively.
[Figure taken from Ref. 26].}
\label{Rxy_pa}
\end{figure}

In a later study, Pryadko and Auerbach\cite{PA} have evaluated the Hall
resistance of a finite quantum coherent CC--network {\em directly}.
They have computed numerically the complex amplitudes associated with
local currents in a network, where the transmission and phase characterizing
coherent scattering at the nodes are random variables. Assuming the network
to be measured by four separate leads labeled $1$ through $4$ (each lead 
connected to a single pair of incoming and outgoing chiral edge channels), 
the Hall voltage for a given total current driven between leads $1$ and $2$
is determined (exactly as in the previous subsection) as the antisymmetric 
component of $V_y(B)=(\mu_3-\mu_4)$. The local chemical potentials $\mu_i$ are 
related to the corresponding local currents through Eq. (\ref{muI}), since 
at the leads a local equilibrium is established. The result of this calculation
are radically different from the classical case [Eq. (\ref{VxyQHI})]: the Hall
resistance in the insulator {\em diverges} exponentially with the system size $L$
(see Fig. \ref{Rxy_pa}). Moreover, $\rho_{xx}$ and $\rho_{xy}$ scale with
each other: $\rho_{xy}\sim\rho_{xx}^\gamma$, where $\gamma\approx 1/3$.    
    
\begin{figure}[b]
\includegraphics[width=2.7in]{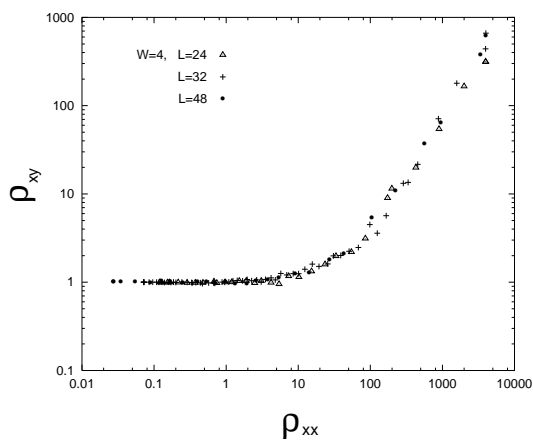}
\caption{$\rho_{xy}$ vs. $\rho_{xx}$ for fixed disorder strength
(denoted by $W$). [Figure taken from Ref. 27].}
\label{Rxy_sw}
\end{figure}

Another numerical study of the coherent transport in a QH system was 
performed by Sheng and Weng\cite{SW} within a tight--binding model
of (non--interacting) electrons in a strong magnetic field at filling
factor $\nu<1$. Using different numerical procedures, they have computed 
the conductivities $\sigma_{xx}$ and $\sigma_{xy}$ and determined the 
resistivities by tensor inversion. Despite significant quantitative 
differences compared to Ref. 26, the insulating regime exhibits the same
behavior: an asymptotic scaling relation correlations between $\rho_{xx}$ and 
$\rho_{xy}$ ($\rho_{xy}\sim\rho_{xx}^\gamma$), both of which diverge 
exponentially with the system size (the exponent $\gamma$ is, however, different 
($\gamma\sim 1$), possibly indicating a model--dependence of its particular value). 
This behavior is demonstrated in Fig. \ref{Rxy_sw} for $\rho_{xx}>10h/e^2$. It 
should be noted, however, that in a wide range of parameters beyond the 
QH--to--Insulator transition (which occurs at $\rho_{xx}^c\approx h/e^2$),
a QHI behavior is indicated. The Hall resistance deviates from the quantized value
$h/e^2$ and starts diverging only deep in the insulating phase, where the system 
size $L$ is sufficiently large compared to the localization length $\xi$ (so that
$\rho_{xx}$ is at least an order of magnitude larger than $\rho_{xx}^c$).

The finite size behavior obtained in the numerical studies described
above effectively simulates a physical macroscopic system 
(where $L\rightarrow\infty$) at finite temperature $T$: in that case 
the finite size cutoff $L$ is replaced by a characteristic
$T$--dependent dephasing length $L_\phi$. The numerical data indicate
that the QHI behavior breaks down at sufficiently low $T$ such that
$L_\phi\gg \xi$, thus raising the intriguing possibility that the breakdown
originates from quantum interference, manifested in terms of localization effects.

\begin{figure}[b]
\includegraphics[width=2.7in]{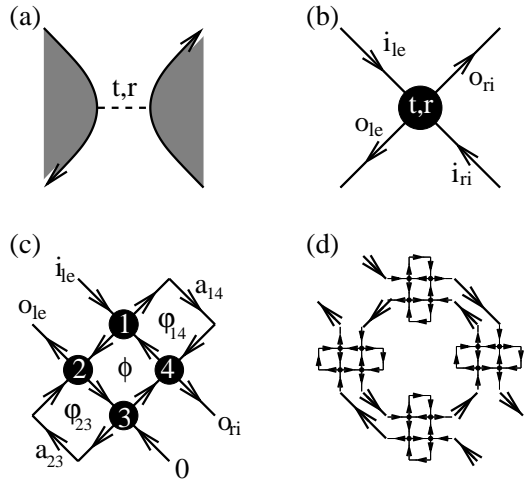}
\vspace{8pt} 
\caption{The hierarchical network model.
(a)~A saddle point between adjacent electron puddles. 
Motion within a puddle is directed along chiral edge channels. 
(b)~Saddle point represented by a scattering matrix (vertex) 
relating outgoing electron amplitudes ($o_{{le}}, 
o_{{ri}}$) to incoming ones ($i_{{le}},i_{{ri}}$). 
(c)~Elementary cell of the hierarchical network. Quantum phases 
acquired by electrons when moving around its three small closed 
loops are indicated. (d)~Lattice at the second level of the 
hierarchy.} 
\label{elemcell} 
\end{figure}

To gain more insight into the underlying localization mechanism, Z{\"u}licke
and Shimshoni\cite{ZS} have studied a model of the QH system which enables the
analytical derivation of scaling expressions for the transport coefficients
$\rho_{xx}$ and $\rho_{xy}$. They consider electron transport on a random 
network that is constructed as a hierarchical lattice (see 
Fig.~\ref{elemcell}). The random variables are the transmission
and reflection amplitudes at the vertices of the elementary cells 
and the phases accumulated in closed loops.

Scaling equations relating the resistivity tensor components to
the number of stages in the hierarchy ($n$) are now derived using a real--space
renormalization group (RG) procedure, which provides a 2D generalization of 
the familiar scattering approach to 1D localization\cite{1Dlocal}. Note that
a similar applications of hierarchical structures have been used to study
the critical behavior near a QH transition\cite{GRAJS}. Here, the focus is
on the asymptotic behavior of transport coefficients deep in the QH and 
insulating phases rather than at the critical point. The procedure is as 
follows: at each RG step, the effective transmission and reflection probability
of a 4--vertex cell in the $n$th level of the hierarchy ($T^{(n)}$, $R^{(n)}$) 
are expressed in terms of $T_i^{(n-1)}$, $R_i^{(n-1)}$ ($i=1,2,3,4$) and the 
phases in the previous level. Then, similarly to Anderson 
{\em et al.}\cite{1Dlocal}, one identifies quantities that are statistically
well--behaved (i.e., their distribution obeys the central limit theorem), so
that their statistical averages over the random variables dictate the 
{\em typical} values of various meaningful physical quantities. In the present
case, the distributions of ln$T^{(n)}$ and ln$R^{(n)}$ are asymptotically 
normal (i.e. for large $n$), and their average values determine the 
scaling expressions for the total current, the Hall and longitudinal voltages
in a system of linear size $L=2^n$. 

\begin{figure}[b]
\includegraphics[width=2.7in]{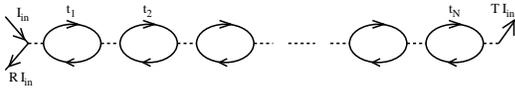}
\vspace*{8pt}
\caption{A serial chain of scatterers in a Hall bar}
\label{serial}
\end{figure}

Before stating the results of this calculation, it is useful to point out that
the study of 1D localization\cite{1Dlocal} already provides much insight on
the expected behavior of a generic QH network. Consider an extreme case of
a very long and narrow Hall bar, where $N$ scatterers are connected serially 
in a chain (Fig. \ref{serial}). According to Ref. 29 the typical transmission
probability of the entire chain, given by 
$T=T_0^N$ where $T_0=\exp\{\langle\ln|t_j|^2\rangle_{t_j}\}$, vanishes 
exponentially with the system size; consequently, the typical longitudinal 
resistance $\rho_{xx}=R/T(h/e^2)$ diverges exponentially:
\begin{equation}\label{serchain}
\rho_{xx}\sim \frac{1}{T_0^N}\frac{h}{e^2}\; .
\end{equation}

Consider now the opposite extreme case of a very short and wide Hall bar, 
where the very same $N$ scatterers are connected {\em in parallel}. This
scenario is also depicted by Fig. \ref{serial} 
provided the total current flows {\em vertically downwards}, 
the coefficients $t_j$ represent the {\em reflection}
amplitudes of the scatterers, and the roles of $T$ and $R$ are exchanged.
It is immediately implied that the system exhibits exponential 
{\em delocalization\/}, i.e. the longitudinal resistance {\em vanishes} as
\begin{equation}\label{parchain}
\rho_{xx}\sim T_0^N\frac{h}{e^2}\; .
\end{equation}
In a general, truly 2D QH network, serial and parallel connections of 
scatterers are completely intertwined and equally abundant. The  
longitudinal transport through a macroscopic system
is therefore expected to exhibit either a perfect 
conducting behavior or an insulating behavior depending on the value of $T_0$.
In particular, a localization--delocalization  transition is expected close to
the symmetric point $T_0=1/2$. We hence deduce that a generic 
QH network in the quantum coherent limit is characterized by a 
``duality symmetry'' between the QH and insulator phases,
of the same origin as the single constriction: it is associated with a 
symmetry to trading of transmission and reflection coefficients combined with
the trading of serial and parallel connections. It is important to emphasize 
that this symmetry property is not related in any way to the behavior of
the Hall resistance in the two phases.

The RG analysis of Ref. 28 indeed accounts for the 2D nature by introducing
a serial and parallel connection at each step. The resulting asymptotic
expression for the typical longitudinal resistivity 
in the limit of large $L=2^n$ is
\begin{equation} 
\rho_{xx}\approx \left(\frac{{\tilde R}_0}{{\tilde T}_0}\right)^L
\frac{h}{e^2}\; ,
\label{longres}
\end{equation}   
where ${\tilde T}_0$ [${\tilde R}_0$] are related\cite{ZS} to the
log--average of the transmission [reflection] coefficients in the
initial distribution (for $n=0$) and its standard deviation. This
indicates a transition from a conducting (QH) state 
($\rho_{xx}\rightarrow 0$) to an insulator ($\rho_{xx}\rightarrow \infty$)
at ${\tilde R}_0={\tilde T}_0$, thus confirming the expectation based on
the naiive extrapolation of Ref. 29. The behavior of the typical Hall 
resistivity is much harder to predict. The calculated 
$\rho_{xy}$ recovers the quantized value in the QH 
phase. However, in the insulator it diverges exponentially:
\begin{equation}\label{hallres} 
\rho_{xy}=\frac{h}{e^2}\left[1-\left({\tilde R}_0\right)^{2^n} 
\right]/\left({\tilde T}_0\right)^{2^{n-1}}\approx 
\frac{h}{e^2}\left({\tilde 
T}_0\right)^{-L/2} \; . 
\end{equation}
Note that similarly to the numerical data\cite{PA,SW},
in the strongly insulating limit (${\tilde T}_0\ll {\tilde R}_0$)
$\rho_{xx}$ and $\rho_{xy}$ obey a scaling relation  
$\rho_{xy}\sim \rho_{xx}^\gamma$ where here $\gamma\approx 1/2$.
The corresponding localization lengths are given by
\begin{equation} 
\xi_{xy}=2\xi_{xx}\; ,\quad {\rm where}\quad \xi_{xx}\approx 
-\frac{1}{\ln {\tilde T}_0}\; .
\end{equation}

The scaling behavior obtained in Ref. 28 supports the 
central conclusion of the numerical studies\cite{PA,SW} --
namely, it indicates a destruction of the QHI by quantum 
coherence effects. Moreover, intermediate stages of the 
analysis\cite{ZS} already reflect a radical difference in 
behavior of the Hall resistance between the QH and insulating phases.
This is in sharp contrast with the classical transport regime. On the 
other hand, the analysis points at certain difficulties which call 
for further investigation. In particular, it is observed that even 
the logarithmic distribution of $\rho_{xy}$ is quite wide and the RG 
equations flow to the asymptotic form quite slowly. In addition, 
the particular structure of the hierarchical network 
(Fig.~\ref{elemcell}), which was chosen to facilitate the analytical
calculation, effectively favors serial connections over parallel ones
in the configuration assumed for the elementary cell. As a result the
duality symmetry is violated: in spite of the ``symmetric appearance" 
of $\rho_{xx}$ [Eq. (\ref{longres})], the critical point for a
transition from the QH phase to the insulator is underestimated.
To see that, note that the universal value $\rho_{xx}=h/e^2$ is
established at ${\tilde R}_0={\tilde T}_0$, however ${\tilde T}_0$
is larger than $T_0$, the typical transmission probability in the
initial distribution of scatterers, implying that the transition occurs 
at $T_0<1/2$. Such a bias is not characteristic of a generic QH system.

A subsequent work by Cain and R{\"o}mer\cite{CR} addresses these problems.
First, they introduce a hierarchical network where the elementary cell
includes a fifth scatterer. The same structure was used earlier by
Galstyan and Raikh\cite{GRAJS} and by Cain {\em et al.}\cite{Cain} in their 
RG studies of the critical exponent, level statistics and long--range
correlations of the disorder potential in the vicinity of the transition.
Apparently, this structure restores the symmetry and yields numerical
values for the critical exponent which are very close to other numerical 
procedures. In addition, a numerical application of the real--space RG 
iterations yields the asymptotics of the entire distributions 
$P(R_{xx})$, $P(R_{xy})$ of both the longitudinal and Hall resistance
components. Identifying $\rho_{xy}$ with either the most probable value 
of $P(R_{xy})$ or the typical value, they observe a QHI behavior up to 
$\rho_{xx}\sim 25h/e^2$. Deeper in the insulator, $\rho_{xy}$ diverges 
and obeys the scaling relation $\rho_{xy}\sim \rho_{xx}^\gamma$ with 
$\gamma=0.26$. Particular details of the crossover between the two
regimes are very sensitive to the averaging procedure.
Note that the numerical value of $\gamma$ is close to the exponent 
found in Ref. 26, possibly reflecting the fact that the network
models studied in the two cases are essentially equivalent.

The theoretical works reviewed above span a variety of different 
models and different approaches to the evaluation of resistivity 
tensors in a macroscopic QH system. Yet, they converge to a
remarkably unique conclusion: that the strictly $T=0$ insulating 
phase in the QH regime is {\em not} a QHI. Rather, quantum 
interference effects induce localization, and divergence of both 
components of the resistivity tensor. The scaling relation 
$\rho_{xy}\sim \rho_{xx}^\gamma$ implies the existence of a single 
localization length $\xi$, to which $\xi_{xx}$ and $\xi_{xy}$ are
related by a numerical factor -- apparently dependent on the geometry
of the disorder potential. In finite size systems, the QHI behavior 
is typically observed in a wide but limited range of parameters 
beyond the critical point, and essentially does not survive in the
thermodynamic limit. The unavoidable conclusion is that in a
realistic system at finite $T$, it does not characterize the 
insulating phase but rather a regime of classical transport,
where $L_\phi<\xi$ thus suppressing quantum interference.  
 
\section{Recent Experimental Results and Open Questions}
The theoretical studies predicting an asymptotic breakdown of the QHI 
also imply that there are severe restrictions on the experimental 
observation of this breakdown. A significant deviation of $\rho_{xy}$ 
from the quantized plateau value is expected in the strong localization 
regime where $L_\phi\gg \xi_{xy}$. Since typically the divergence of 
$\rho_{xy}$ is relatively moderate ($\xi_{xy}>\xi_{xx}$), it appears 
that this quantum regime is approached only when the longitudinal 
resistivity $\rho_{xx}$ is more than an order of magnitude larger than 
its critical value of $\sim h/e^2$. The appropriate conditions can be
achieved at very low $T$ and strong magnetic fields, far enough from
the critical value $B_c$. In addition, better conditions for localization
are established in highly disordered samples. This regime of parameters
is accessible; however, the main problem is that it is hard to obtain 
reliable experimental data deep in the insulating phase. In particular,
due to the wide distribution of the Hall resistance\cite{CR}, its 
measured value is expected to be very sensitive to the measurement
procedure and the external circuitry. 

A recent experimental study\cite{Buth} has focused on magnetotransport
measurements in GaAs/GaAlAs heterostructures that are deliberately 
contaminated by Be acceptors in a $\delta$--doping layer. These 
introduce a short range disorder, hence expected to reduce the 
localization length in the insulating state. Preliminary data obtained
at low $T$ (down to $T=60$mK) indicate a considerable deviation of
$\rho_{xy}$ from the quantized value $h/e^2$ when $\rho_{xx}$ increases 
beyond 100K$\Omega$. Deeper in the insulating regime, where 
$\rho_{xx}>10h/e^2$, a divergence of $\rho_{xy}$ (up to $\sim 20h/e^2$) 
is observed, and the scaling relation $\rho_{xy}\sim \rho_{xx}^\gamma$ 
is approximately obeyed. Due to the subtleties of the measurement setup
the error bars on $\rho_{xy}$ and hence on the numerical value of
$\gamma$ are relatively large ($\gamma$ is estimated in the range $0.5$ 
to $0.75$), however the divergence of $\rho_{xy}$ far beyond the 
quantized value appears to be conclusive.

Yet another recent experimental result\cite{Peled} can possibly be 
interpreted as a challenge to the theoretical predictions in the
quantum coherent limit. Magnetotransport measurements near the
transition from a $\nu=1$ QH state to the insulator were performed
on small samples of low--mobility InGaAs/InAlAs (of a few microns
in each dimension) at low $T$. The trace of $\rho_{xx}$ as a 
function of the magnetic field $B$ was found to display strong 
reproducible fluctuations, whose amplitude increases when 
the sample size is reduced. In contrast, $\rho_{xy}$ in the same
range of parameters is quantized to a very good approximation.
Since the fluctuations appear to be highly reminiscent of 
universal conductance fluctuations (UCF) that are known to 
reflect the quantum coherence nature of transport in mesoscopic 
samples, this observation suggests that quantum interference 
effects are present yet the QHI behavior is not destroyed. 
Rather, the fluctuations in the component $\sigma_{xx}$,
$\sigma_{xy}$ of the conductivity tensor are correlated in such
a way that their fluctuations conspire to cancel each other in 
the expression for $\rho_{xy}$. A similar correlation was
also identified near the transition between QH 
plateaus\cite{Peled2}.

There are two possible interpretations for these experimental 
results. The first option assumes that (even at the low $T$ 
tested in these measurements) $L_\phi$ is sufficiently small to 
impose the conditions of classical transport in the system.
In that case, $\rho_{xy}$ is obviously quantized, and 
$\rho_{xx}$ is dictated by the classically connected network
of resistors associated with junctions between QH electron puddles.
The observed fluctuations in $\rho_{xx}$ are then a consequence
of the statistical distribution of $\log[R_{xx}]$ in a small
system where the number of ``resistors" ($N\approx L/l_{el}$,
where $L$ is the linear size of the system and $l_{el}$ the elastic
mean free path) is far from the thermodynamic limit. In the present case,
$l_{el}\sim 0.1\mu$m implying that in the smallest measured samples
$\sqrt{N}\approx 20$. Indeed, the average size of the fluctuations is
consistent with the expected standard deviation. However, the spiky
nature of the $\rho_{xx}$ trace as a function of $B$ is not consistent
with the $B$--dependence expected from variations in the width of 
tunnel--barriers between QH puddle. The rapid fluctuations can be 
alternatively attributed to charging effects involved in the hopping
processes between QH puddles, a mechanism suggested following earlier 
experimental observations of mesoscopic fluctuations\cite{Cobden}.

The second possibility is that the fluctuations indeed reflect the
effect of a magnetic field on the interference pattern between different
electron paths in a quantum coherent transport regime. This mechanism
could result in reproducible fluctuations of a similar character as UCF,
although their amplitude should not be necessarily universal (more likely,
it is dictated by the typical value of the resistivity). According to the
theories described in subsection 3.3, such interpretation is apparently in 
conflict with the QHI behavior. 

It should be noted, however, that the 
present theoretical understanding is capable of definite predictions in either
of two cases. In the extreme ``classical transport" limit, where $L_\phi$ is 
smaller than a puddle size, interference is entirely suppressed and 
$\rho_{xy}$ is quantized\cite{SA}. In the extreme ``quantum transport" limit, 
where $L_\phi$ is larger than the localization length $\xi$, quantum 
interference destroys the quantization in the insulator and leads to a 
divergence of $\rho_{xy}$ in the thermodynamic limit\cite{PA,SW,ZS,CR}. 
The puddle size is of order $l_{el}$, and since close to the transition 
$\xi$ is typically much larger than $l_{el}$ there is an intermediate range 
where the theory is practically not decisive about 
the behavior of $\rho_{xy}$ (especially in view of its wide distribution).
The experimental data\cite{Peled}, which mostly accumulate within this
intermediate range (where $\rho_{xx}$ increases by at most one order of 
magnitude), possibly provide evidence for the stability of the QHI behavior in 
the entire region where $\xi>L_\phi$. 

A better theoretical understanding of the transport behavior in the above
mentioned intermediate regime would require a detailed study of the mechanism
leading to destruction of phase coherence, and correspondingly a concrete 
evaluation of the dephasing length $L_\phi$. In section 3 it was 
essentially defined as the typical length scale over which the interference
between different electron trajectories is suppressed. Dephasing is 
provided by coupling of the system to other degrees of freedom and by 
electron--electron interactions, hence $L_\phi$ generally depends on many 
details such as the strength of various coupling constants. However, it is
possible that close enough to the transition and at sufficiently low $T$, 
some of the dephasing processes become irrelevant and the dynamics is 
dictated by universal properties, signifying the quantum critical nature
of the transition point\cite{qhe-phas-tran,SGCS}. $L_\phi$ should then 
be identified with the correlation time $\xi_T\sim \xi^z$ (with $z$ the
dynamical exponent), and the crossover line $\xi\sim L_\phi$ (which
apparently marks the breakdown of the QHI) is to be identified with the
borderline of the quantum critical regime\cite{Sachdev}. Unfortunately, a
convenient field--theoretical model which enables a direct analysis of
the transport properties at finite $T$ within this framework is presently
lacking.

On the experimental front, further investigation is required to distinguish
conclusively between the different possible interpretations of the 
fluctuations in mesoscopic samples. For example, the study of the effect of
a screening gate could test the significance of Coulomb blockade; a Fourier
analysis of the $B$--dependence of the fluctuations in $\rho_{xx}$, and the 
study of correlations between the typical periodicity in magnetic flux and 
the scale $l_{el}$, may shed light on the role of quantum interference.   

\section{Conclusions}
As demonstrated in a multitude of experiments reviewed in this article, the
insulating regime neighboring a primary QH liquid state $\nu=1/k$ (with $k$ 
an odd integer) exhibits a QHI behavior: a quantized Hall resistance 
accompanied by an insulating--like character of the longitudinal transport.
The phenomenon has been studied theoretically using various models, and in
particular network models for the transport in the QH regime. The results of
these studies indicate that the QHI does not characterize the full--fledged 
insulating phase, which is established sufficiently far from the transition.
Nor is there any evidence for an additional phase transition (from a QHI to
a ``true'' insulator) within the insulating regime. Rather, it is predicted
that in the $T=0$ insulating phase, where transport is dominated by quantum 
interference, the Hall resistance should diverge. The QHI behavior is
characteristic of a ``classical'' transport regime, where quantum coherence 
on a scale larger then $l_{el}$ (a typical size of a QH puddle) is suppressed 
by a dephasing mechanism. Transport in this regime is therefore best modeled
as a resistor network, where the individual resistors exhibit quantum features,
yet the connections between them obey the {\em classical} Kirchoff's laws.
Preliminary experimental data support the breakdown of the QHI deep in the 
insulator, where localization on length scales larger than $\xi$ takes place. 
There is, however, recent evidence for a robust quantization of the Hall 
resistance in the entire regime where $\xi$ is of order or larger than the 
dephasing length $L_{\phi}$. It is suggestive that this regime can be 
identified as a quantum critical regime, yet a full understanding of the 
transport mechanism requires further investigation.

Finally, a comment is in order regarding the initial observation of a 
connection between the QHI behavior and duality symmetry\cite{SSSexp,SSSthe}. 
Later studies certainly support the suggestion that in the quantum critical 
regime both features coincide. However, in view of the network models of
the transport in the QH and neighboring insulator phase, it appears that
the two phenomena are actually distinct. Duality symmetry is associated
with the symmetry (on average) to curvature inversion of the saddle--points
in the disorder potential (and thus to trading transmission and reflection
coefficients), and does not depend on the transport being quantum coherent
or classical. In contrast, the lack of quantum coherence and the localization
effects derived from it is crucial for the QHI phenomenon. Duality symmetry
in the longitudinal transport is possible even in a regime where the 
Hall resistance in the insulator is not quantized. In principle, in samples 
where particle--hole symmetry around the critical filling factor is strongly
violated, duality symmetry may not be obeyed, yet the QHI behavior is 
established as long as quantum interference in the global transport is 
suppressed. This leads to an interesting conjecture about the set of arguments
previously leading to the conclusion that the two features are interwined 
within the composite bosons picture (see Section 2) -- 
e.g., the application of the ``resistivity law" [Eq. (\ref{reslaw})]. 
It is possible that they actually rely on an implicit assumption that 
the transport properties are classical, Boltzmann--like in nature. 

\bigskip

{\leftline{\bf acknowledgements}}

The author wishes to thank A. Auerbach, D. Shahar, M. Shayegan, S. L. Sondhi,
D. C. Tsui and U. Z\"ulicke for the fruitful collaborative research on the
topics reviewed in this article. She is also greatful to these esteemed 
colleagues, as well as to A. Aharony, D. P. Arovas, K. Buth, S. Chakravarty, 
J.~T. Chalker, N. Cooper, O. Entin-Wohlman, H. A. Fertig, M. Fogler, 
E. Fradkin, Y. Gefen, S. M. Girvin, B. Halperin, M. Hilke, Y. Imry, 
J. K. Jain, A. Kapitulnik, D. Khmelnitskii, Y. Meir, F. von Oppen, E. Peled, 
L.~P. Pryadko, M.~E. Raikh, R. A. R{\"o}mer, S. Sachdev, B. Shapiro, 
D.~N. Sheng, S. Simon and A. Stern for numerous illuminating discussions.
A. Auerbach, M. Hilke, L.~P. Pryadko and D.~N. Sheng are greatfully 
acknowledged for providing figures 2, 5 and 6 in this article. Finally,
the author thanks O. Entin-Wohlman, M. Hilke, L. P. Pryadko, D. Shahar
and U. Z\"ulicke for their careful review of the manuscript and 
constructive comments.


\begin{thebibliography}{0}
\bibitem{Fuku}
H. Fukuyama, {\it J. Phys. Soc. Jpn.} {\bf 49}, 644 (1980); B. Altshuler, 
D. Khmelnitskii, A. Larkin and P. A. Lee, {\it Phys. Rev.}  {\bf B22}, 
5142 (1980).
\bibitem{Efetov}
O. Viehweger and K. B. Efetov, {\it Phys. Rev.}  {\bf B44}, 1168 (1991). 

\bibitem{KLZlett}
S. C. Zhang, S. Kivelson and D. H. Lee, {\it Phys. Rev. Lett.} {\bf 69}, 
1252 (1992).

\bibitem{Joe}
Y. Imry, {\it Phys. Rev. Lett.} {\bf 71}, 1868 (1993).

\bibitem{Herb}
Many body effects where considered by
L. Zheng and H. A. Fertig, {\it Phys. Rev. Lett.} {\bf 73}, 878 (1994);
L. Zheng and H. A. Fertig, {\it Phys. Rev.} {\bf B50}, 4984 (1994).

\bibitem{Ora}
O. Entin-Wohlman, A. G. Aronov, Y. Levinson and Y. Imry, {\it Phys. Rev. 
Lett.} {\bf 75}, 4094 (1995); see also Comment by O. Bleibaum, 
H. B{\"o}ttger and V. V. Bryksin, {\it Phys. Rev. Lett.} 
{\bf 79}, 2752 (1997) and subsequent Reply.

\bibitem{FrPo}
L. Friedman and M. Pollak, {\it Philos. Mag.} {\bf B44}, 487 (1981).

\bibitem{Hop}
P. Hopkins, M. J. Burns, A. J. Rimberg and R. M. Westervelt, {\it Phys. Rev.} 
{\bf B39}, 12708 (1989). 

\bibitem{qhe-phas-tran} 
For reviews and extensive lists of references, see A.~M.~M. 
Pruisken, in {\it The Quantum Hall Effect}, 2nd ed., edited by R.~E. Prange 
and S.~M. Girvin (Springer, NY, 1990); S. Das Sarma in {\it
Perspectives in Quantum Hall Effects}, edited by S. Das Sarma and 
A. Pinczuk (Wiley, NY, 1997); B. Huckestein, {\it Rev. Mod. 
Phys.} {\bf 67},  357 (1995).
\bibitem{KLZ}
S. Kivelson, D. H. Lee and S. C. Zhang, {\it Phys. Rev.} {\bf B46}, 
2223 (1992).
\bibitem{Vladimir}
V. J. Goldman, M. Shayegan and D. C. Tsui, {\it Phys. Rev. Lett.} {\bf 61}, 
881 (1988); V. J. Goldman, J. K. Wang, B. Su and M. Shayegan, {\it Phys. Rev. 
Lett.} {\bf 70}, 647 (1993).
\bibitem{Willett}
R. L. Willett, H. L. Stormer, D. C. Tsui, L. N. Pfeiffer,
K. W. West and K. W. Baldwin, {\it Phys. Rev.} {\bf B38}, 7881 (1988).
\bibitem{Shahar1}
D. Shahar, D. C. Tsui, M. Shayegan, R. N. Bhatt and J. E. Cunningham,
{\it Phys. Rev. Lett.} {\bf 74}, 1511 (1995).
\bibitem{SSSexp}
D. Shahar, D. C. Tsui, M. Shayegan, E. Shimshoni and S. L. Sondhi, 
{\it Science} {\bf 274}, 589 (1996).
\bibitem{SSSthe}
E. Shimshoni, S. L. Sondhi and D. Shahar, {\it Phys. Rev.} {\bf B55},
13730 (1997).
\bibitem{qhi-expA} 
D. Shahar, D. C. Tsui, M. Shayegan, J. E. Cunningham, E. Shimshoni 
and S. L. Sondhi, {\it Solid State Comm.} {\bf 102}, 817 (1997).
\bibitem{qhi-expB}
M. Hilke, D. Shahar, S. H. Song, D. C. Tsui, Y. H. Xie and Don Monroe, 
{\it Nature}  {\bf 395}, 675 (1998).
\bibitem{qhi-expC} 
M.~V. Yakunin, Yu. G. Arapov, O. A. Kuznetsov and V. N. Neverov, 
{\it JETP Lett.} {\bf 70}, 301 (1999); D. T. N. de Lang, L. A. Ponomarenko,
A. de Visser, C. Possanzini, S. M. Olsthoorn and A. M. M. Pruisken,
{\it Physica E} {\bf 12}, 666 (2002).
\bibitem{Trugman}
A. Trugman, {\it Phys. Rev.} {\bf B27}, 7539 (1983).
\bibitem{LanBut}
R. Landauer, {\it Philos. Mag.} {\bf 21}, 863 (1970); M. B{\"u}ttiker,
Y. Imry, R. Landauer and S. Pinhas, {\it Phys. Rev.} {\bf B31}, 6207 (1985).

\bibitem{JK} 
J. K. Jain and S. A. Kivelson, {\it Phys. Rev.} {\bf B37}, 4276 (1988).    

\bibitem{Been}
C.W.J. Beenakker and H. van Hoiuten, {\it Solid State Physics: 
Advances in Research and Applications}, Ed. H. Ehrenreich and D. Turnbull 
(Academic, San Diego, 1991), Vol 44, pp. 207, 208.

\bibitem{CC}
J.~T. Chalker and P.~D. Coddington, {\it J. Phys.} {\bf C21},  2665  
(1988). 

\bibitem{ruzin}
A. M. Dykhne and I. M. Ruzin, {\it Phys. Rev.} {\bf B50}, 2369 (1994);
I. M. Ruzin and S. Feng, {\it Phys. Rev. Lett.} {\bf 74}, 154 (1995).

\bibitem{SA}
E. Shimshoni and A. Auerbach, {\it Phys. Rev.} {\bf B55}, 9817 (1997).

\bibitem{PA} 
L.~P. Pryadko and A. Auerbach, {\it Phys. Rev. Lett.} {\bf 82},  1253  
(1999). 

\bibitem{SW}
D.~N. Sheng and Z.~Y. Weng, {\it Phys. Rev.} {\bf B59},  R7821 
(1999). 

\bibitem{ZS}
U. Z\"ulicke and E. Shimshoni, {\it Phys. Rev.} {\bf B63}, R241301 (2001).
 
\bibitem{1Dlocal} 
P.~W. Anderson, D.~J. Thouless, E. Abrahams, and D.~S. Fisher, 
{\it Phys. Rev.} {\bf B22},  3519  (1980).

\bibitem{GRAJS}
A.~G. Galstyan and M.~E. Raikh, {\it Phys. Rev.} {\bf B56},  1422  
(1997); D.~P. Arovas, M. Janssen, and B. Shapiro, {\it Phys. Rev.} {\bf B56},
4751 (1997).

\bibitem{CR}
P. Cain and R. A. R{\"o}mer, {\it Europhys. Lett.} {\bf 66}, 104 (2004).

\bibitem{Cain}
P. Cain, R. A. R{\"o}mer, M. Schreiber and M.~E. Raikh, {\it Phys. Rev.} 
{\bf B64}, 235326 (2001); P. Cain, R. A. R{\"o}mer and M.~E. Raikh, 
{\it Phys. Rev.} {\bf B67}, 075307 (2003). 

\bibitem{Buth}
K. Buth {\em et al.}, unpublished.

\bibitem{Peled}
E. Peled, D. Shahar, Y. Chen, D. L. Sivco and A. Y. Cho, {\it Phys. Rev.
Lett.} {\bf 90}, 246802 (2003).

\bibitem{Peled2}
E. Peled, D. Shahar, Y. Chen, E. Diez, D. L. Sivco and A. Y. Cho, 
{\it Phys. Rev. Lett.} {\bf 91}, 236802 (2003).

\bibitem{Cobden}
D. H. Cobden, C. H. W. Barnes and C. J. B. Ford, {\it Phys. Rev. Lett.} 
{\bf 82}, 4695 (1999). 

\bibitem{SGCS}
S. L. Sondhi, S. M. Girvin, J. P. Carini and D. Shahar, {\it Rev. Mod. Phys.}
{\bf 69}, 315 (1997).

\bibitem{Sachdev}
S. Sachdev, {\it Quantum Phase Transitions} (Cambridge University Press,
Cambridge, 1999).

\end{thebibliography}
\end{document}